\documentclass[12pt]{article}
\usepackage{amsmath,amssymb}
\usepackage{graphicx}

\numberwithin{equation}{section}
\numberwithin{figure}{section}
\numberwithin{table}{section}

\begin{document}

\begin{titlepage}
	\begin{flushright}
		UT-Komaba/12-13
	\end{flushright}
	\begin{center}

		\vspace*{15mm}

		{\LARGE\bf Detectability of the second resonance\\[2mm]of low-scale string models at the LHC}
 
		\vspace*{20mm}

		{\large Manami Hashi$\,^a$ and Noriaki Kitazawa$\,^b$}

		\vspace{6mm}

		$^a\,${\it Institute of Physics, University of Tokyo,\\
		Komaba 3-8-1, Meguro-ku, Tokyo 153-8902, Japan\\
		e-mail: hashi@hep1.c.u-tokyo.ac.jp}\\

		\vspace{3mm}

		$^b\,${\it Department of Physics, Tokyo Metropolitan University,\\
		Hachioji, Tokyo 192-0397, Japan\\
		e-mail: kitazawa@phys.se.tmu.ac.jp}

		\vspace*{15mm}

		\begin{abstract}

Low-scale string models are phenomenological models in String Theory, 
in which the string scale $M_\mathrm{s}$ is of the order of TeV.
String excited states which are characteristic modes in low-scale string models 
can be observed as resonances in dijet invariant mass distributions at the LHC.
If a new heavy resonance is discovered at the LHC, 
it is important to investigate whether the resonance comes from low-scale string models.
In this work, two analyses are performed:
One is observing higher spin degeneracy of string excited states 
by an angular distribution analysis on the resonance, 
since the string resonance consists of several degenerate states with different spins.
The other is observing second string excited states 
by a search for a second resonance in dijet invariant mass distributions, 
since second string excited states have characteristic masses of $\sqrt{2}$ times of masses of first string excited states.
As the result of Monte Carlo simulations assuming the $14\,\mathrm{TeV}$ LHC, 
we give required luminosities for $5\,\sigma$ confirmation in each analysis, 
in case of $M_\mathrm{s}=4.5$, $4.75$ and $5\,\mathrm{TeV}$.

		\end{abstract}

	\end{center}
\end{titlepage}

\newpage

\section{Introduction}
	\label{sec:introduction}

Low-scale string models are phenomenological models in String Theory, 
 whose fundamental scale, the string scale $M_\mathrm{s}$, 
 is of the order of TeV \cite{Antoniadis:1990ew,Antoniadis:1998ig}.
Since such a low string scale is possible due to the existence of large extra dimensions, 
 these models are expected to solve the hierarchy problem of the Standard Model (SM) 
 and have a possibility of being confirmed or excluded by the Large Hadron Collider (LHC) \cite{Cullen:2000ef}.
Signatures of low-scale string models at the LHC are different 
from other typical signatures of field theoretical models with extra dimensions, 
such as the ADD model \cite{ArkaniHamed:1998rs} and the RS model \cite{Randall:1999ee}. 

A characteristic feature in low-scale string models is the appearance of string excited states.
Since string excited states have the same gauge quantum number of the corresponding SM particle,
 colored string excited states can be produced in dijet events at the LHC \cite{Anchordoqui:2008di,Anchordoqui:2009mm}.
They can be observed as resonances with mass of $M_\mathrm{s}$ in dijet invariant mass distributions.\footnote{
 It is pointed out in Ref.\cite{Hassanain:2009at}, for example, that light string excited states are possible 
 even with a high string scale due to the warp effect in the Randall-Sundrum scenario.
 }
Scattering amplitudes with exchanges of string excited states are calculated using string world-sheet theory, 
 and a parameter in the amplitudes is the string scale $M_\mathrm{s}$ only.
The value of $M_\mathrm{s}$ has already been constrained by dijet events at the $7\,\mathrm{TeV}$ LHC, 
 to be larger than $4.31\,\mathrm{TeV}$ by the CMS with $5.0\,\mathrm{fb}^{-1}$ \cite{CMS:2012yf}, 
 and $3.61\,\mathrm{TeV}$ by the ATLAS with $4.8\,\mathrm{fb}^{-1}$ \cite{ATLAS:2012pu}.

String excited states correspond to vibrational modes of string and these have the following unique properties.
There are several states degenerate in mass with a variety of spins higher than the spin of the corresponding SM particle.
Their degenerate masses are $M_n=\sqrt{n}M_\mathrm{s}$ 
 and their highest spins are $J_{max}=j_0+n$ for $n$th string excited states, 
 where $j_0$ is the spin of the corresponding SM particle.
For example, first string excited states of gluons, $g^*$s, are degenerate with spin $J=0,\,1$ and $2$, 
 and first string excited states of quarks, $q^*$s, are degenerate with $J=1/2$ and $3/2$.
All of them have the same mass of $M_\mathrm{s}$.
In two-parton scattering processes at the LHC, 
 dominant processes are $gg\rightarrow gg$ and $gg\rightarrow q\bar{q}$ 
 where the $g^*$s can be produced in $s$-channel.
Another one is $qg\rightarrow qg$ where the $q^*$s can be produced.
These processes do not contain exchanges of Kaluza-Klein (KK) modes of SM particles 
 because of momentum conservation in the directions of extra dimensions.
They are independent of the detail of the model buildings, 
 such as the ways of compactifications of extra dimensions, 
 assuming the ``local model'' \cite{Lust:2008qc} in which 
 the geometry of compact space is general and SM branes are wrapped with local cycles 
 (see Ref.\cite{Hashi:2011cz} for details).

If a new heavy resonance is discovered at the LHC, 
 it is important to distinguish whether the resonance comes from low-scale string models or the other ``new physics''.
In this work, two analyses are performed in a similar way to the previous works \cite{Hashi:2011cz,Kitazawa:2010gh}.
One is observing degeneracy of string excited states with higher spins 
 by an angular distribution analysis on the resonance in dijet invariant mass distributions.
The other is observing second string excited states with characteristic masses 
 by a search for a second resonance in dijet invariant mass distributions.
In order to perform these analyses, we run Monte Carlo (MC) simulations for the $14\,\mathrm{TeV}$ LHC.

The process of $qg\rightarrow qg$ almost dominates over all the other processes 
 due to the effect of distribution functions of quarks with large momentum fractions.
Therefore, the $q^*$s which have not only $J=1/2$ but also $J=3/2$ are dominant in the string resonance.
Analysing dijet angular distributions on the string resonance, 
 we show that the angular distribution is fitted for that with both $J=1/2$ and $3/2$ states 
 better than that with a $J=1/2$ state only, 
 with a certain given luminosity.
In the previous work \cite{Hashi:2011cz}, a reference value of the string scale $M_\mathrm{s}$ was $4\,\mathrm{TeV}$, 
 however, in this work, we take the values of $4.5\,\mathrm{TeV}$ and $5\,\mathrm{TeV}$.

There are second string excited states in low-scale string models, 
 while there is no second state in the other ``new physics'' 
 such as axigluon models \cite{Frampton:1987dn} and color-octet scalar models \cite{Han:2010rf}.
The masses of second string excited states are $\sqrt{2}M_\mathrm{s}$, 
 or $\sqrt{2}$ times of that of first string excited states, 
 while typical masses of second KK modes in the other ``new physics'' with extra dimensions 
 are 2 times of that of first KK modes.
We show that the second string resonance can be discovered at $5\,\sigma$ level at the $14\,\mathrm{TeV}$ LHC,
 for $M_\mathrm{s}=4.5$, $4.75$ and $5\,\mathrm{TeV}$, 
 assuming that second string excited states do not decay into first string excited states.
We give a simple estimate of decay widths of second string excited states including the above decay processes, 
 and see that the High-Luminosity LHC is needed to discover the second string resonance.

In Sec.\ref{sec:string_amplitude} string amplitudes for dijet events are reviewed. 
In Sec.\ref{sec:angular_analysis} the angular analysis of dijet events is given, 
 and in Sec.\ref{sec:second_resonance} the second resonance analysis in dijet invariant mass distributions is given. 
Finally, in Sec.\ref{sec:conclusion} our results are summarized.

\section{String amplitudes for dijet events}
	\label{sec:string_amplitude}

In low-scale string models, two-body scattering amplitudes between SM particles are calculated 
 as open-string amplitudes with four external legs of lowest vibrational modes of open string.
The open-string amplitudes reduce to the SM amplitudes in the low-energy limit 
 where a scattering energy $\sqrt{s}$ is much less than the string scale $M_\mathrm{s}$, $\sqrt{s}\ll M_\mathrm{s}$.
The amplitudes begin to represent string effects at $\sqrt{s}\gtrsim M_\mathrm{s}$, 
 through ``string form factor'' functions such as
\begin{equation}
	\label{eq:V-function}
	V(s,t,u)
	= \frac{\Gamma(1-s/M_\mathrm{s}^2)\Gamma(1-u/M_\mathrm{s}^2)}{\Gamma(1+t/M_\mathrm{s}^2)} \,,
\end{equation}
 where $s$, $t$ and $u$ are the Mandelstam variables of the scattering SM particles.
The form factor function is expanded by a sum over infinite $s$-channel poles,
\begin{equation}
	\label{eq:V-function_expansion}
	V(s,t,u)
	\simeq \sum^\infty_{n=1} \frac{1}{(n-1)!} \frac{1}{(M_\mathrm{s}^2)^{n-1}} \frac{1}{s-nM_\mathrm{s}^2}
	\prod_{J=0}^{n-1} (u+JM_\mathrm{s}^2) \,,
\end{equation}
 which is a good approximation near each $n$th pole, $s\simeq nM_\mathrm{s}^2$.
These poles correspond to string excited states which have masses of $M_n=\sqrt{n}M_\mathrm{s}$.
This shows that the four-point open-string amplitudes reduce 
 to the two-body scattering amplitudes between SM particles, 
 in which string excited states are exchanged in $s$-channel.

Angular dependences in the open-string amplitudes are described by a factor 
 $\prod_{J=0}^{n-1}(u+JM_\mathrm{s}^2)$ in eq.(\ref{eq:V-function_expansion}) 
 multiplied by angular dependences in the original SM amplitudes.
The open-string amplitudes are decomposed into a sum over the Wigner $d$-functions, 
 $d^J_{J_z,J_{z^\prime}}(\theta)$.\footnote{The Wigner $d$-function $d^J_{J_z,J_{z^\prime}}(\theta)$ represents 
  an angular dependence of a process through a state with spin $J$, 
  in which an initial state has total spin along $z$-axis, $J_z$, 
  and a final state has total spin along $z^\prime$-axis, $J_{z^\prime}$. 
 The angle $\theta$ is that between the $z$-axis and $z^\prime$-axis.}
This shows that $n$th string excited states corresponding to the $n$th pole in eq.(\ref{eq:V-function_expansion}) 
 are degenerate in mass of $\sqrt{n}M_\mathrm{s}$ with various spins of $J$ 
 which are smaller than a highest spin $J_{max}$.
The value of the highest spin is $J_{max}=j_0+n$, where $j_0$ is the spin of the original SM particle.

String excited states decay into SM particles, and the $s$-channel poles in eq.(\ref{eq:V-function_expansion}) 
 are modified to the Breit-Wigner form
\begin{equation}
	\label{eq:Breit-Wigner_form}
	\frac{1}{s-M_n^2}
	\rightarrow \frac{1}{s-M_n^2+iM_n\Gamma_n^J}\,,
\end{equation}
 where $\Gamma^J_n$ is a total decay width of the $n$th string excited state with spin $J$.
The width is calculated by extracting couplings among the string excited state and SM particles 
 from coefficients of each Wigner $d$-function $d^J_{J_z,J_{z^\prime}}(\theta)$ 
 in the corresponding open-string amplitudes \cite{Anchordoqui:2008hi}.

The dominant process at the LHC is $qg\rightarrow qg$ due to the effect of distribution functions of $q$, 
 where $q$ denotes $u$-quark or $d$-quark.
In the process of $qg\rightarrow qg$, first string excited states of quarks, $q^*$s, 
 can be produced at $\sqrt{s}\simeq M_\mathrm{s}$, 
 and second string excited states of quarks, $q^{**}$s, 
 can be produced at $\sqrt{s}\simeq\sqrt{2}M_\mathrm{s}$.
The $q^*$s are degenerate two states with $J=1/2$ and $3/2$, 
 and the $q^{**}$s are degenerate three states with $J=1/2$, $3/2$ and $5/2$.

The spin- and color-averaged squared amplitude of $qg\rightarrow qg$ 
 with exchanges of $q^*$s is calculated 
 in Ref.\cite{Anchordoqui:2008di,Anchordoqui:2009mm,Anchordoqui:2008hi},\footnote{The complete squared amplitude 
  of $qg\rightarrow qg$ is given 
  by a sum of eq.(\ref{eq:1st_squared_amplitude_qg_t-channel}) 
  and that with a replacement of $\hat{u}\leftrightarrow\hat{t}$ in eq.(\ref{eq:1st_squared_amplitude_qg_t-channel}).}
\begin{equation}
	\begin{split}
		\label{eq:1st_squared_amplitude_qg_t-channel}
	 \bigl| \mathcal{M}_\mathrm{1st} ( qg \rightarrow qg ) \bigr|^2
	 = & \frac{N^2-1}{2N^2}
 	 \frac{g_s^4}{M_\mathrm{s}^2}
	 \Biggl[
	 	\frac{ M_\mathrm{s}^4 (-\hat{u}) }
		         { ( \hat{s} - M_\mathrm{s}^2 )^2
		         + \bigl( M_\mathrm{s} \Gamma_{q^*}^{J=1/2} \bigr)^2 }
	    + \frac{ (-\hat{u})^3 }
	                { ( \hat{s} - M_\mathrm{s}^2 )^2
	                + \bigl( M_\mathrm{s} \Gamma_{q^*}^{J=3/2} \bigr)^2 }
	 \Biggr] \,,
	\end{split}
\end{equation}
 where $N=3$, $g_s$ is the gauge coupling constant of strong interaction, 
 and $\hat{s}$, $\hat{t}$ and $\hat{u}$ are the Mandelstam variables of the scattering quark and gluon. 
Here, $\Gamma_{q^*}^J$s are decay widths of the $q^*$s with spin $J$ 
 due to a decay process of $q^*\rightarrow qg$.
They are explicitly given as
\begin{equation}
	\label{eq:1st_quark_width}
	 \Gamma_{q^*}^{J=1/2}
	 = \frac{g_s^2}{4\pi}
	    M_\mathrm{s}
	    \frac{N}{8} \,,
	    \hspace{6mm}
	 \Gamma_{q^*}^{J=3/2}
	 = \frac{g_s^2}{4\pi}
	    M_\mathrm{s}
	    \frac{N}{16} \,,
\end{equation}
 for $J=1/2$ and $3/2$, respectively.

On the other hand, the squared amplitude of $qg\rightarrow qg$ with exchanges of $q^{**}$s 
 is calculated in Ref.\cite{Hashi:2011cz},
\begin{equation}
	\begin{split}
		\label{eq:2nd_squared_amplitude_t-channel}
	 & \bigl| \mathcal{M}_\mathrm{2nd} ( qg \rightarrow qg ) \bigr|^2 \\
	 = & \frac{2(N^2-1)}{N^2}
 	 \Biggl\{
		 \frac{g_s^4}{2M_\mathrm{s}^2}
		 \Biggl[
		 	\hspace{2mm}
		 	\frac{1}{9}
			\frac{ M_\mathrm{s}^4 ( -\hat{u} ) }
			        { \bigl( \hat{s} - 2M_\mathrm{s}^2 \bigr)^2
			       + \bigl( \sqrt{2}M_\mathrm{s} \Gamma^{J=1/2}_{q^{**}} \bigr)^2 }
			\hspace{1mm}
		   + \hspace{1mm}
			\frac{1}{9}
			\frac{ ( -\hat{u} ) ( 3\hat{t} + \hat{s} )^2 }
			         { \bigl( \hat{s} - 2M_\mathrm{s}^2 \bigr)^2
			        + \bigl( \sqrt{2}M_\mathrm{s} \Gamma^{J=3/2}_{q^{**}} \bigr)^2 }
		\Biggr] \\
	    & \hspace{17mm}
	    + \frac{g_s^4}{8M_\mathrm{s}^6}
	       \Biggl[
	    		\frac{9}{25}
			\frac{ M_\mathrm{s}^4 ( -\hat{u} )^3}
			         { \bigl( \hat{s} - 2M_\mathrm{s}^2 \bigr)^2
			        + \bigl( \sqrt{2}M_\mathrm{s} \Gamma^{J=3/2}_{q^{**}} \bigr)^2 }
		    + \frac{1}{25}
		       \frac{ ( -\hat{u} )^3 ( 5\hat{t} + \hat{s} )^2 }
		                { \bigl( \hat{s} - 2M_\mathrm{s}^2 \bigr)^2
		               + \bigl( \sqrt{2}M_\mathrm{s} \Gamma^{J=5/2}_{q^{**}} \bigr)^2 }
		 \Biggr]
	\Biggr\} \,,
	\end{split}
\end{equation}
 where $\Gamma_{q^{**}}^J$s are decay widths of the $q^{**}$s with spin $J$ 
 due to a decay process of $q^{**}\rightarrow qg$.
They are explicitly given as
\begin{equation}
	\label{eq:2nd_quark_width}
	 \Gamma^{J=1/2}_{q^{**}}
	 = \frac{g_s^2}{4\pi}
	    \sqrt{2}M_\mathrm{s}
	    \frac{N}{24} \,,
	    \hspace{6mm}
	 \Gamma^{J=3/2}_{q^{**}}
	 = \frac{g_s^2}{4\pi}
	    \sqrt{2}M_\mathrm{s}
	    \frac{19N}{240} \,,
	    \hspace{6mm}
	 \Gamma^{J=5/2}_{q^{**}}
	 = \frac{g_s^2}{4\pi}
	    \sqrt{2}M_\mathrm{s}
	    \frac{N}{60} \,,
\end{equation}
 for $J=1/2$, $3/2$ and $5/2$, respectively.
Note that interference effects are neglected in eq.(\ref{eq:2nd_squared_amplitude_t-channel}). 

Now consider interference effects between first and second string excited states 
 and between second string excited states.
An amplitude with exchanges of $q^*$s and $q^{**}$s with $J_z=\pm1/2$ 
 (for example, in processes of $q^\pm g^\pm\rightarrow q^\pm g^\pm$) 
 is written as a sum of the amplitudes with exchanges of the $q^{*}$ with $J=1/2$, 
 the $q^{**}$ with $J=1/2$ and the $q^{**}$ with $J=3/2$ :
\begin{equation}
	\label{eq:1st_and_2nd_amplitude_Jz=1/2}
	\mathcal{M}_{1\mathrm{st}+2\mathrm{nd}}^{J_z=\pm1/2}
	= \mathcal{M}_{1\mathrm{st}}^{J=1/2}
	+ \mathcal{M}_{2\mathrm{nd}}^{J=1/2}
	+ \mathcal{M}_{2\mathrm{nd}}^{J=3/2} \,.
\end{equation}
The squared amplitude of the process includes three interference terms
\begin{equation}
	\label{eq:1st_and_2nd_interference_Jz=1/2}
	\mathcal{M}_{1\mathrm{st}}^{J=1/2} {\mathcal{M}_{2\mathrm{nd}}^{J=1/2}}^*
	+ \mathcal{M}_{1\mathrm{st}}^{J=1/2} {\mathcal{M}_{2\mathrm{nd}}^{J=3/2}}^*
	+ \mathcal{M}_{2\mathrm{nd}}^{J=1/2}{\mathcal{M}_{2\mathrm{nd}}^{J=3/2}}^* \,,
\end{equation}
 and their complex conjugations. 
The last term in eq.(\ref{eq:1st_and_2nd_interference_Jz=1/2}), 
 an interference between second string excited states,
 gives large contributions to the total squared amplitude at $\sqrt{s}\simeq\sqrt{2}M_\mathrm{s}$.
The other two terms in eq.(\ref{eq:1st_and_2nd_interference_Jz=1/2}) give small suppression effects 
 in the region of $M_\mathrm{s}<\sqrt{s}<\sqrt{2}M_\mathrm{s}$.
Since open-string amplitudes with the expansion of eq.(\ref{eq:V-function_expansion}) 
 are good approximations only near $s$-channel poles, 
 such as $s\simeq M_\mathrm{s}^2$ and $s\simeq2M_\mathrm{s}^2$, 
 the inclusion of the first two terms in eq.(\ref{eq:1st_and_2nd_interference_Jz=1/2}) may make this analysis imprecise.
In this work, however, we focus on only the neighborhood of the $s$-channel poles, 
 and the effect of these terms are negligible.

In the decay widths of $q^{**}$s in eq.(\ref{eq:2nd_quark_width}), 
 only the decay process of $q^{**}\rightarrow qg$ is considered, 
 since they are calculated by extracting couplings 
 from the two-body scattering amplitude of $qg\rightarrow q^{**}\rightarrow qg$. 
The masses of second string excited states, $\sqrt{2}M_\mathrm{s}$, 
 are larger than that of first string excited states, $M_\mathrm{s}$, 
 and decay processes of $q^{**}\rightarrow q^*g$ and $q^{**}\rightarrow qg^*$ should also be considered.
However, there are some technical complication 
 to calculate string amplitudes with external legs of first string excited states.
We discuss this issue in Sec.\ref{sec:second_resonance}.

\section{Angular analyses of dijet events}
	\label{sec:angular_analysis}

If a new heavy resonance is discovered at the LHC, 
 an angular analysis is important to confirm that the resonance comes from low-scale string models.
In the dominant process at the LHC, $qg\rightarrow qg$, 
 there are two degenerate string excited states of quarks with $J=1/2$ and $3/2$.
If we can experimentally distinguish angular distributions
 with both $J=1/2$ and $3/2$ or with $J=1/2$ only, 
 it can be a signature of low-scale string models.

In this work, since we consider two-parton scattering processes,
 performing computer simulations for hadronization and detector simulation is required.
The setup of our MC simulations is as follows.
First, we use $\mathtt{CalcHEP}$ \cite{Pukhov:1999gg,Pukhov} to generate event samples at parton-level.
We generate event samples for the SM background of two-parton scattering processes at tree-level, 
 as well as event samples for the string signal of $qg\rightarrow qg$ 
 by including the string amplitudes 
 such as eqs.(\ref{eq:1st_squared_amplitude_qg_t-channel}) and (\ref{eq:2nd_squared_amplitude_t-channel}) 
 in the program (see the web page of \cite{Kitazawa}).
Next, we use $\mathtt{Pythia8}$ \cite{Sjostrand:2007gs} for hadronization of quarks and gluons in the final state, 
 and use $\mathtt{Delphes1.9}$ \cite{Ovyn:2009tx} for detector simulation 
 mainly to identify jets consisting of hadrons using its default detector card for the ATLAS.
Finally, we use $\mathtt{ROOT}$ \cite{Brun:1997pa} to construct event samples of dijet, 
 by choosing two energetic jets, $j_1$ and $j_2$, 
 with largest and second largest transverse momenta event by event.
We also use $\mathtt{ROOT}$ for suppressing background events 
 and relatively enhancing signal events by imposing cuts on various kinematical observables in dijet events.

We analyse $\chi$-distributions as angular distributions of dijet events, 
 which is used by the ATLAS to search ``new physics'' beyond the SM \cite{Aad:2011aj}.
The quantity $\chi$ is defined as
\begin{equation}
	\label{eq:chi}
	\chi
	= \exp{(y_1-y_2)}
	= \frac{1+\cos\theta_*}{1-\cos\theta_*} \,,
\end{equation}
 where $y_1$ and $y_2$ are pseudo rapidities of two jets, 
 and $\theta_*$ is a scattering angle in the parton center-of-mass frame.

The formula of the $\chi$-distribution on the first string resonance is derived in the following way.
A differential cross section is described 
 by a spin-averaged squared amplitude as $d\sigma/d\hat{t}=\bigl|\mathcal{M}\bigr|^2/16\pi^2\hat{s}$.
By substituting the squared amplitude of eq.(\ref{eq:1st_squared_amplitude_qg_t-channel}), 
 a prediction to the $\chi$-distribution with first string excited states of quarks, $q^*$s, 
 with $J=1/2$ and $3/2$ is obtained as
\begin{equation}
	\label{eq:chi_distribution}
	\frac{d\sigma_\mathrm{1st}(qg\rightarrow qg)}{d\chi}
	= \frac{1}{(1+\chi)^2} \biggl(C^{J=1/2}+C^{J=3/2}\frac{1+\chi^3}{(1+\chi)^3}\biggr)\,,
\end{equation}
 where $C^{J=1/2}$ and $C^{J=3/2}$ are constants.
The term proportional to $C^{J=1/2}$ represents a $\chi$-dependence of the $q^*$ with $J=1/2$ 
 and the term proportional to $C^{J=3/2}$ represents that of the $q^*$ with $J=3/2$.
The overall factor $1/(1+\chi)^2$ in eq.(\ref{eq:chi_distribution}) is a kinematical factor.

The kinematical cuts which are imposed on both background and signal event samples in this analysis are
\begin{equation}
	\begin{split}
		\label{eq:angular_analysis_cut}
	p_{\mathrm{T},j_1} > 400\,\mathrm{GeV} \hspace{2mm}
	\mathrm{for} \hspace{2mm} M_\mathrm{s} = 4.5\,\mathrm{TeV} & \,, \hspace{4mm}
	p_{\mathrm{T},j_1} > 450\,\mathrm{GeV} \hspace{2mm}
	\mathrm{for} \hspace{2mm} M_\mathrm{s} = 5\,\mathrm{TeV} \,, \\
	|y_1-y_2| < 2.3 & \,, \hspace{4mm}
	|y_1+y_2| < 2.0 \,,
	\end{split}
\end{equation}
 assuming the LHC with the center-of-mass energy $14\,\mathrm{TeV}$.
To obtain events near the first string resonance, 
 we use only events in the dijet invariant mass window of 
 $M_{jj}=[M_\mathrm{s}-250\,\mathrm{GeV},\,M_\mathrm{s}+250\,\mathrm{GeV}]$.
Two event samples for the SM background and for the SM background plus string signal are generated, 
 and signal event samples for this analysis are obtained by substracting the former from the latter.

\begin{center}
	\begin{figure}[b]
	\begin{minipage}{0.325\hsize}
		\centering
		\includegraphics[width=62mm]
		{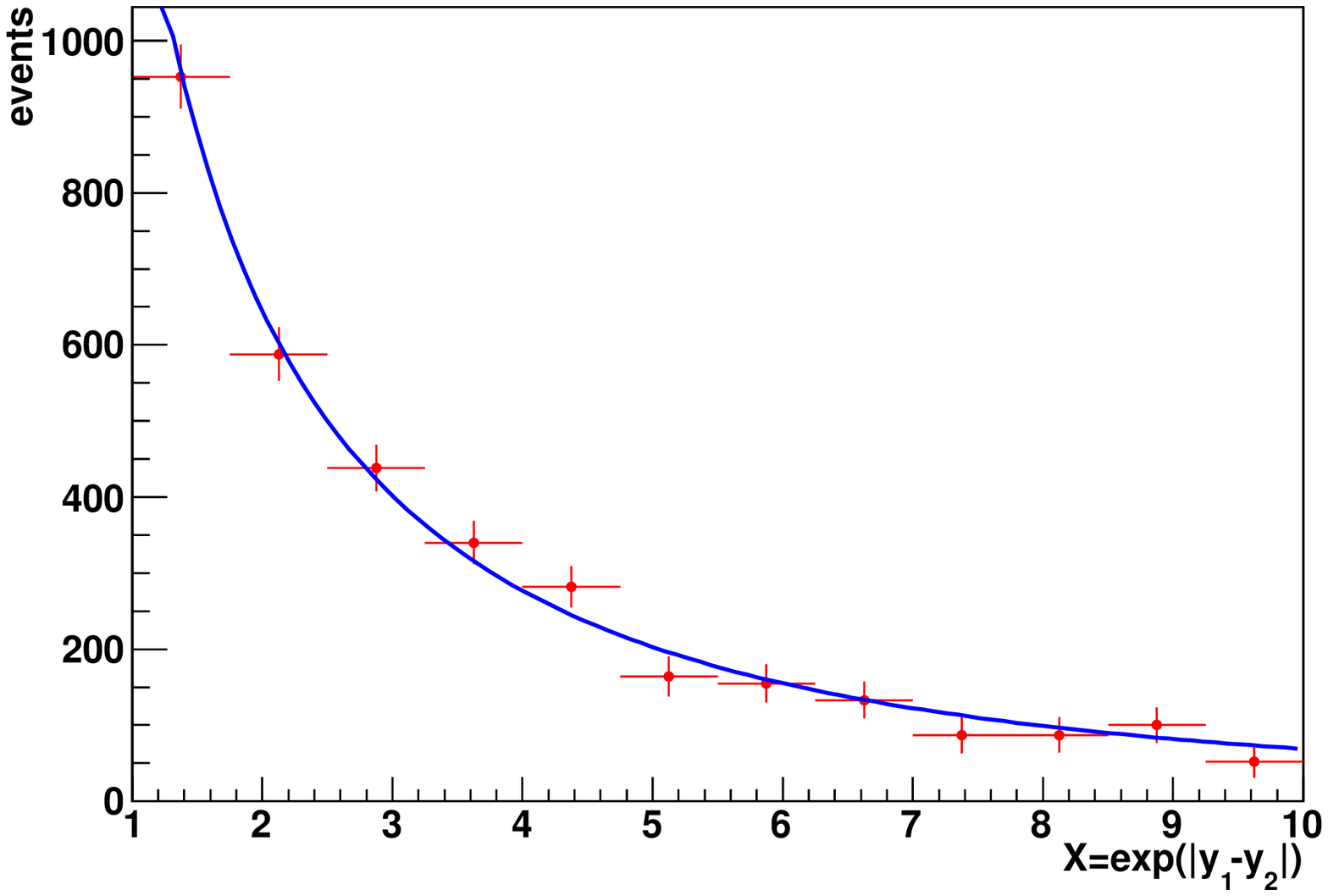}
	\end{minipage}
	\begin{minipage}{0.325\hsize}
		\centering
		\includegraphics[width=62mm]
		{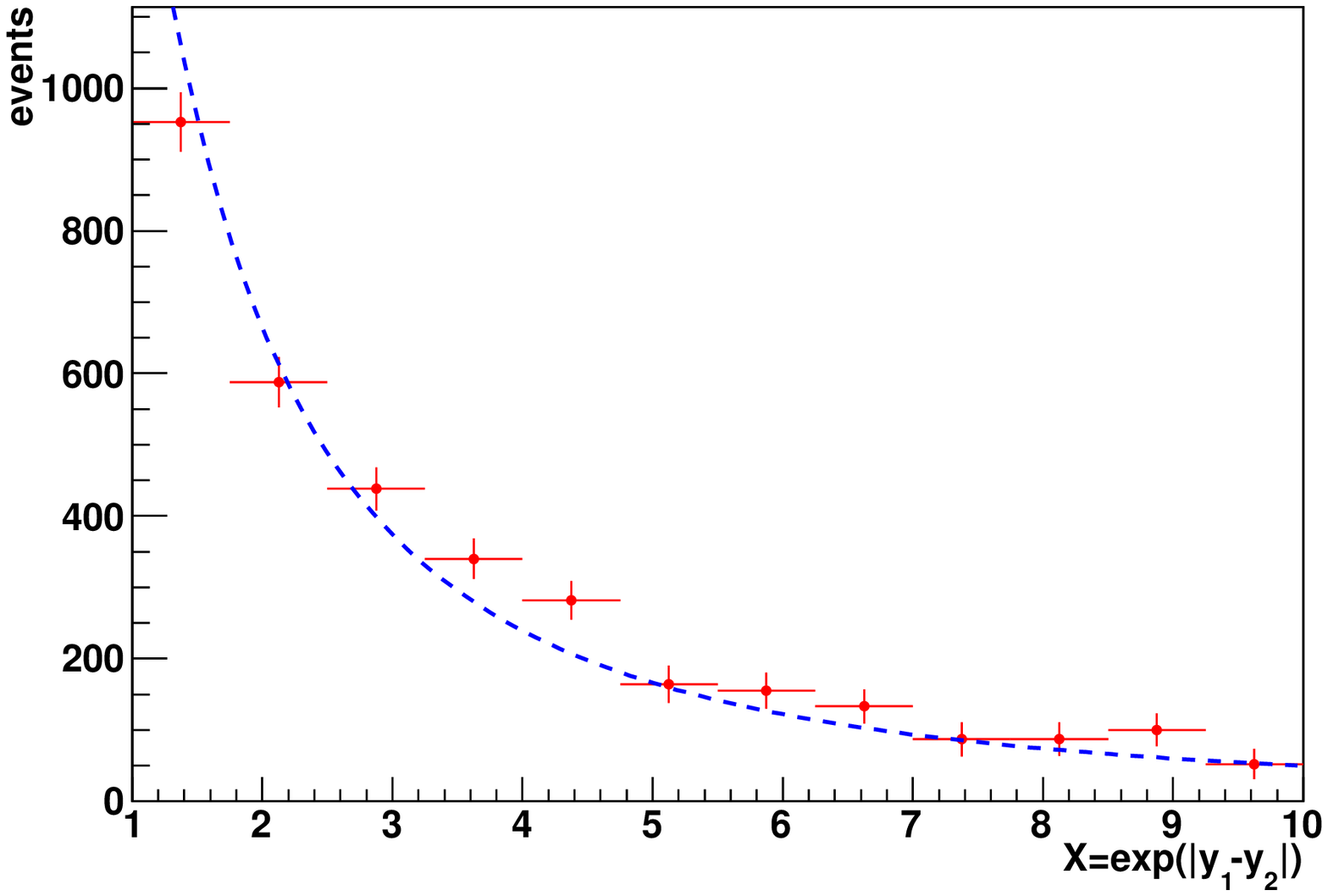}
	\end{minipage}
	\begin{minipage}{0.325\hsize}
		\centering
		\includegraphics[width=62mm]
		{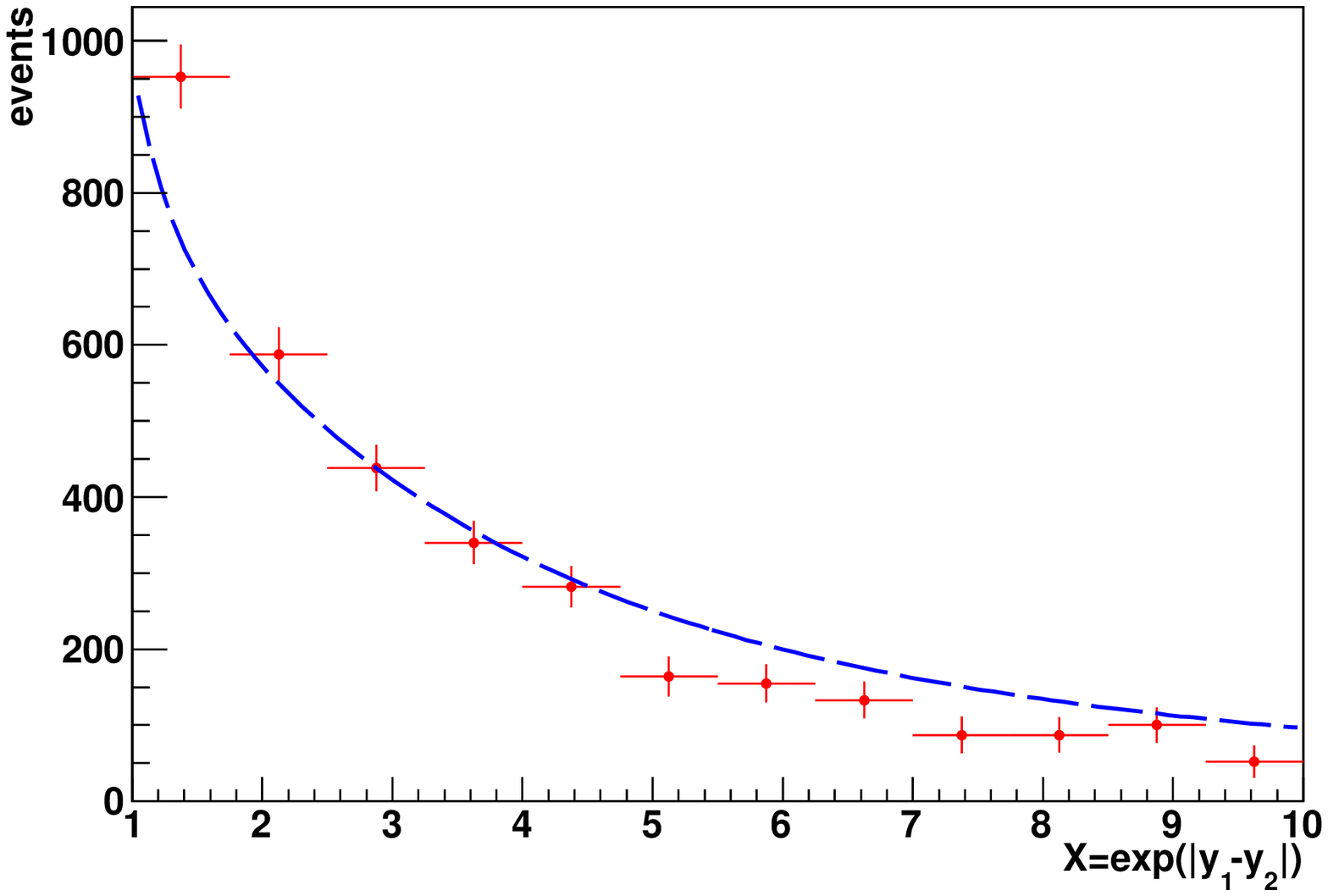}
	\end{minipage}
	\caption{\small{The $\chi$-distributions of MC data 
	for $M_\mathrm{s}=4.5\,\mathrm{TeV}$ with $30\,\mathrm{fb}^{-1}$ (the points with error bars), 
	and fits of three hypotheses with both $J=1/2$ and $3/2$ (the solid line in the left figure), 
	$J=1/2$ only (the dashed line in the middle figure) 
	and $J=3/2$ only (the long dashed line in the right figure).}}
	\label{fig:chi_4.5TeV}
	\end{figure}
\end{center}

\begin{center}
	\begin{figure}[t]
	\begin{minipage}{0.325\hsize}
		\centering
		\includegraphics[width=62mm]
		{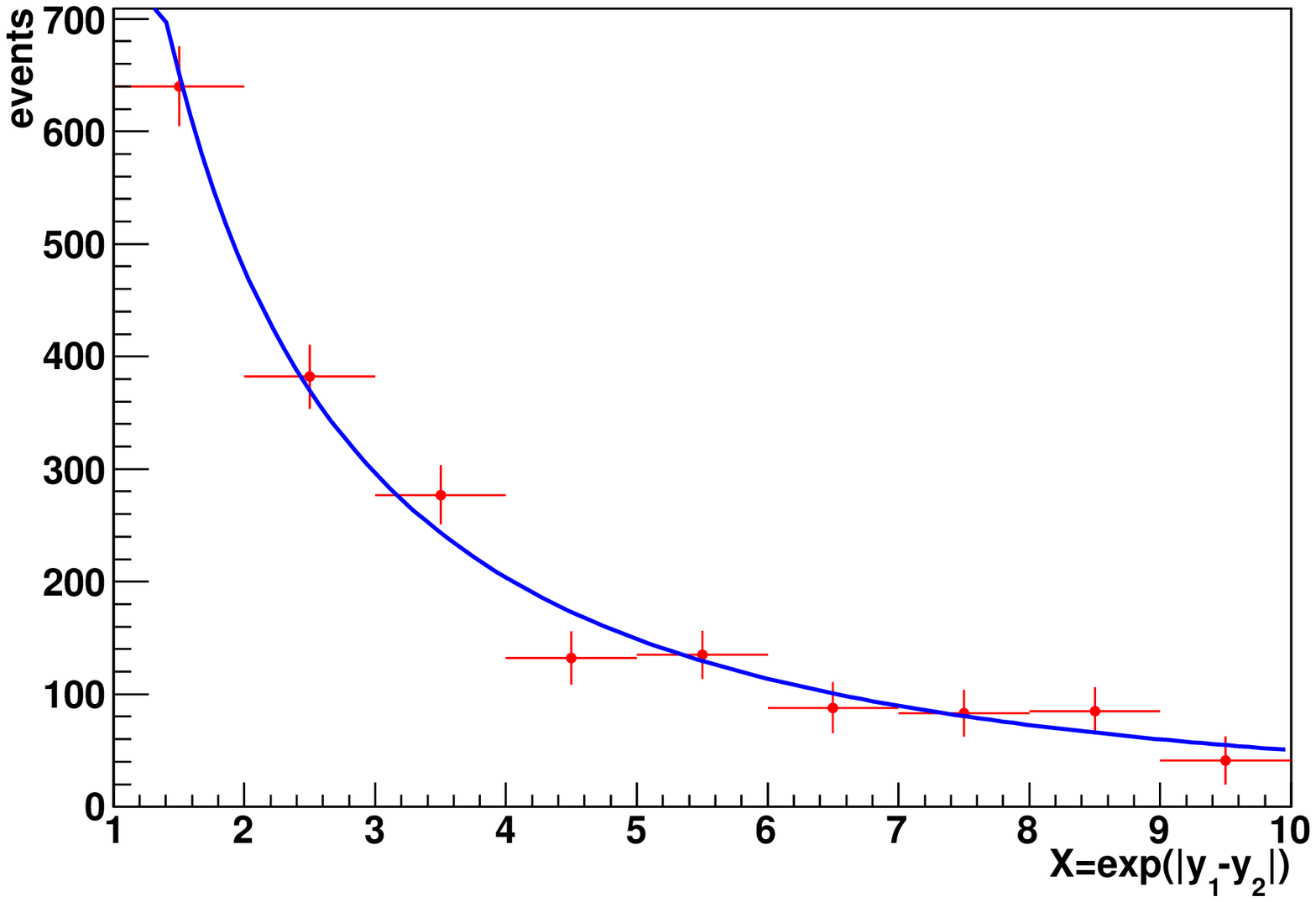}
	\end{minipage}
	\begin{minipage}{0.325\hsize}
		\centering
		\includegraphics[width=62mm]
		{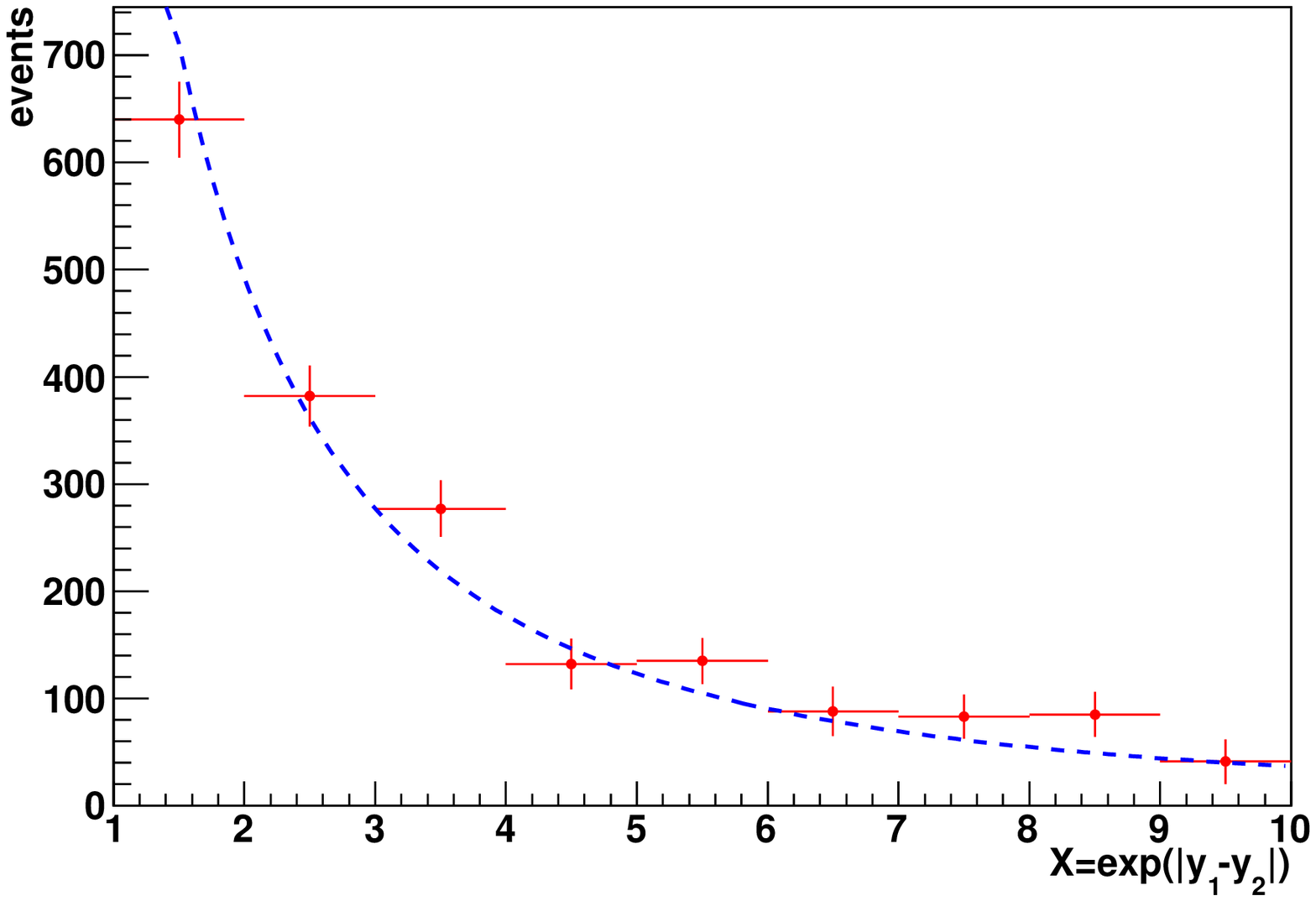}
	\end{minipage}
	\begin{minipage}{0.325\hsize}
		\centering
		\includegraphics[width=62mm]
		{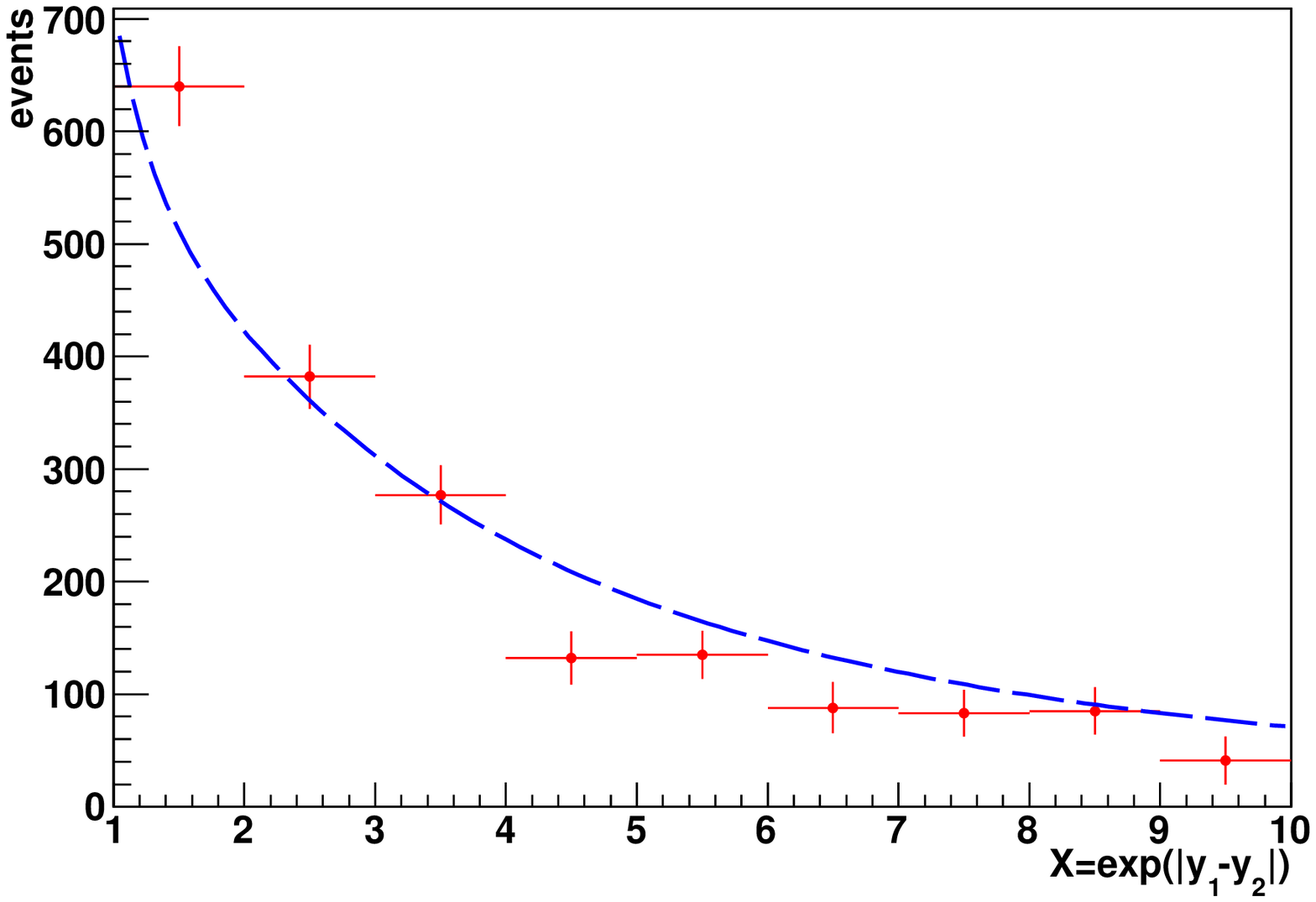}
	\end{minipage}
	\caption{\small{The $\chi$-distributions of MC data 
	for $M_\mathrm{s}=5\,\mathrm{TeV}$ with $50\,\mathrm{fb}^{-1}$ (the points with error bars), 
	and fits of three hypotheses with both $J=1/2$ and $3/2$ (the solid line in the left figure), 
	$J=1/2$ only (the dashed line in the middle figure) and $J=3/2$ only (the long dashed line in the right figure).}}
	\label{fig:chi_5TeV}
	\end{figure}
\end{center}

We fit $\chi$-distributions of signal event samples using functions of $\chi$ in eq.(\ref{eq:chi_distribution}), 
 in three cases of $C^{J=1/2}\neq0$ and $C^{J=3/2}\neq0$, 
 $C^{J=1/2}\neq0$ and $C^{J=3/2}=0$, 
 and $C^{J=1/2}=0$ and $C^{J=3/2}\neq0$.
These three cases correspond to assumptions with both $J=1/2$ and $3/2$ states, 
 a $J=1/2$ state only, and a $J=3/2$ state only, respectively.
The first assumption corresponds to low-scale string models with $q^*$s, 
 while the second one corresponds to the other ``new physics'' with new quark-like particles.
Figs.\ref{fig:chi_4.5TeV} and \ref{fig:chi_5TeV} show the $\chi$-distributions of signal event samples 
 and these three fits, 
 for the cases of $M_\mathrm{s}=4.5\,\mathrm{TeV}$ and $5\,\mathrm{TeV}$ 
 with $30\,\mathrm{fb}^{-1}$ and $50\,\mathrm{fb}^{-1}$ of integrated luminosities, respectively, 
 at the $14\,\mathrm{TeV}$ LHC.

\vspace{-22mm}
Since the $\chi$-distributions clearly do not match the fits with $J=3/2$ only 
 in the right figures of Figs.\ref{fig:chi_4.5TeV} and \ref{fig:chi_5TeV}, 
 the hypothesis with $J=3/2$ only can be excluded.
On the other hand, other two hypotheses with both $J=1/2$ and $3/2$ in the left figures 
 and with $J=1/2$ only in the middle figures of Figs.\ref{fig:chi_4.5TeV} and \ref{fig:chi_5TeV} 
 give rather good fits.
It is difficult to distinguish whether there are two states with $J=1/2$ and $3/2$ 
 or there is only a state with $J=1/2$ in the resonance, 
 without a statistical analysis.
We give $p$-values of these fits in Tab.\ref{tab:p-value}.

\begin{table}[h]
	\begin{center}
	\vspace{4mm}
	\begin{tabular}{|c|c|c|c|}\hline
	$M_\mathrm{s}\,(\mathrm{luminosity})$       & $J=1/2$ and $3/2$
	 & $J=1/2$ only             & $J=3/2$ only                                   \\ \hline
	$4.5\,\mathrm{TeV}\,(30\,\mathrm{fb}^{-1})$   & $0.6966$
	 & $0.004562$ & $5.386\times10^{-9}$ \\
	$5\,\mathrm{TeV}\,(50\,\mathrm{fb}^{-1})$ & $0.4737$
	 & $0.0466$   & $3.198\times10^{-5}$                                   \\ \hline
	\end{tabular}
	\end{center}
	\vspace{-3mm}
	\caption{$p$-values of the fits}
	\label{tab:p-value}
\end{table}

The $p$-value is a probability that if a hypothesis is excluded in spite of that it is correct, 
 the exclusion is an experimental error.\footnote{See G. Cowan, 
  \textit{``Statistics,''} on page 320-329 of \cite{Amsler:2008zzb}.}
If the $p$-value is smaller than $5\%$, the hypothesis can be excluded at $95\%$ confidence level.
Evidently, the $p$-values of the hypotheses with $J=1/2$ only and with $J=3/2$ only 
 are very small and smaller than $5\%$ in Tab.\ref{tab:p-value}.
On the other hand, the $p$-value of the hypothesis with both $J=1/2$ and $3/2$ is much larger than $5\%$, 
 and the hypothesis cannot be excluded, 
 for $M_\mathrm{s}=4.5\,\mathrm{TeV}$ with $30\,\mathrm{fb}^{-1}$ 
 and $M_\mathrm{s}=5\,\mathrm{TeV}$ with $50\,\mathrm{fb}^{-1}$.
If we look at the middle figures of Figs.\ref{fig:chi_4.5TeV} and \ref{fig:chi_5TeV} closer, 
 it is found that the fit with $J=1/2$ only is systematically inconsistent, 
 since curves of the fit falls more quickly than the MC data for large $\chi$.

The spin degeneracy of string excited states 
 can be experimentally confirmed with $\mathcal{O}(10)\,\mathrm{fb}^{-1}$ of integrated luminosity 
 by the angular analysis on a new resonance in dijet invariant mass distributions.
This can be a signature of low-scale string models.

\section{Second resonance in dijet invariant mass distributions}
	\label{sec:second_resonance}

To confirm that a new resonance comes from low-scale string models, 
 a search for a second resonance in dijet invariant mass distributions is important.
The existence of second string excited states and their characteristic masses 
 are distinct properties of low-scale string models.
The masses of second string excited states are $\sqrt{2}$ times of that of first string excited states.
On the other hand, typical masses of second KK modes are $2$ times of that of first KK modes 
 in models with KK modes of SM particles, 
 such as the five-dimensional Universal Extra Dimension (UED) models.\footnote{In the six-dimensional UED models, 
 typical masses of second KK modes with KK parity $+1$ 
  is $\sqrt{2}$ times of that of first KK modes with KK parity $+1$. 
  It may be possible to confirm low-scale string models by a search for the third resonance, 
  since third string excited states have $\sqrt{3}$ times of that of first string excited states.}
In this analysis, we investigate the potential of discovery of the second string resonance in dijet invariant mass distributions.

Dijet event samples are obtained using the same setup of MC simulations described in Sec.\ref{sec:angular_analysis}.
The kinematical cuts in this analysis are
\begin{equation}
	\begin{split}
		\label{eq:second_resonance_analysis_cut}
	p_{\mathrm{T},j_1} > 400\,\mathrm{GeV} & \hspace{2mm}
	\mathrm{for} \hspace{2mm} M_\mathrm{s} = 4.5\,\mathrm{TeV} \,, \\
	p_{\mathrm{T},j_1} > 430\,\mathrm{GeV} & \hspace{2mm}
	\mathrm{for} \hspace{2mm} M_\mathrm{s} = 4.75\,\mathrm{TeV} \,, \\
	p_{\mathrm{T},j_1} > 450\,\mathrm{GeV} & \hspace{2mm}
	\mathrm{for} \hspace{2mm} M_\mathrm{s} = 5\,\mathrm{TeV} \,, \\
	|y_1|, \, |y_2| < 2.3 \,, & \hspace{4mm}
	|y_1-y_2| < 1.7 \,.
	\end{split}
\end{equation}
Event samples for the SM background plus string signal are generated as experimental data 
 assuming that low-scale string models are realized, 
 while event samples for the SM background are generated as theoretical predictions of the SM.

\begin{figure}[t]
	\centering
	\includegraphics[width=62mm]{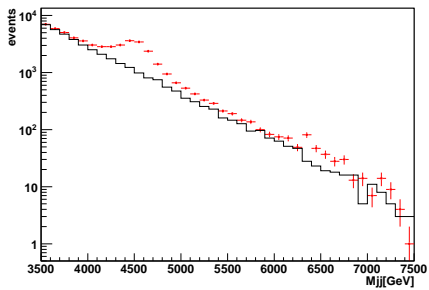}
	\caption{\small{The dijet invariant mass distribution of MC data 
	with the first and second string resonance for $M_\mathrm{s}=4.5\,\mathrm{TeV}$ (the points with error bars) 
	and of the SM background (the histogram) with $100\,\mathrm{fb}^{-1}$ at the $14\,\mathrm{TeV}$ LHC.}}
	\label{fig:1st_and_2nd_4.5TeV}
\end{figure}

The dijet invariant mass distribution with the first and second string resonances 
 is shown in Fig.\ref{fig:1st_and_2nd_4.5TeV}, 
 for $M_\mathrm{s}=4.5\,\mathrm{TeV}$, 
 assuming that $100\,\mathrm{fb}^{-1}$ of integrated luminosity is given at the $14\,\mathrm{TeV}$ LHC.
We can see a first resonance at $M_{jj}=M_\mathrm{s}=4.5\,\mathrm{TeV}$ 
 and a second resonance at $M_{jj}=\sqrt{2}\times M_\mathrm{s}\simeq6.36\,\mathrm{TeV}$.

To investigate the potential of discovery of the second string resonance, 
 we calculate a signal significance which describes a deviation from the SM prediction 
 within the dijet invariant mass window of 
 $M_{jj}=[\sqrt{2}M_\mathrm{s}-250\,\mathrm{GeV},\,\sqrt{2}M_\mathrm{s}+250\,\mathrm{GeV}]$.
If the signal significance is larger than $5\,\sigma$, 
 we claim that a second resonance can be discovered.

The significance  $Z$ is calculated in the following way.
We define a statistic $\chi^2$ as follows
\begin{equation}
	\label{eq:chi_square}
	\chi^2
	= \sum_{i=1}^{N_\mathrm{bin}}
	\frac{\bigl(\#\,\mathrm{of}\,\mathrm{events}\,\mathrm{for}\,\mathrm{SM}\,\mathrm{plus}\,\mathrm{string}\,-\,\#\,\mathrm{of}\,\mathrm{events}\,\mathrm{for}\,\mathrm{SM}\bigr)^2}{\#\,\mathrm{of}\,\mathrm{events}\,\mathrm{for}\,\mathrm{SM}\,\mathrm{plus}\,\mathrm{string}\,+\,\#\,\mathrm{of}\,\mathrm{events}\,\mathrm{for}\,\mathrm{SM}\hspace{1.5mm}} \,,
\end{equation}
 where $N_\mathrm{bin}$ is the number of bins.
The number of events for the SM plus string and that for the SM in eq.(\ref{eq:chi_square}) 
 are the numbers of events in a bin $i$ of the dijet invariant mass.
The numerator represents a deviation of ``experimental results'' from the SM hypothesis.
The denominator represents a sum of statistical dispersions 
 of ``experiment'' and of the SM prediction, 
 $\sigma_\mathrm{exp.}^2+\sigma_\mathrm{MC}^2$, 
 for each bin (assuming that the number of events follows Poisson distribution).
Since the statistic $\chi^2$ follows $\chi^2$ distribution if the SM hypothesis is correct, 
 the $p$-value for $\chi^2$ is calculated by $p=\int_{\chi^2}^\infty f(z;N_\mathrm{bin})dz$, 
 where $f(z;N)$ is the probability distribution function of $\chi^2$ distribution with $N$ number of degree of freedom.
Then, the significance is calculated as $Z=\Phi^{-1}(1-p)$ 
 where $\Phi$ is the cumulative distribution function of standard Gaussian distribution.

\begin{center}
	\begin{figure}[t]
	\begin{minipage}{0.325\hsize}
		\centering
		\includegraphics[width=62mm]{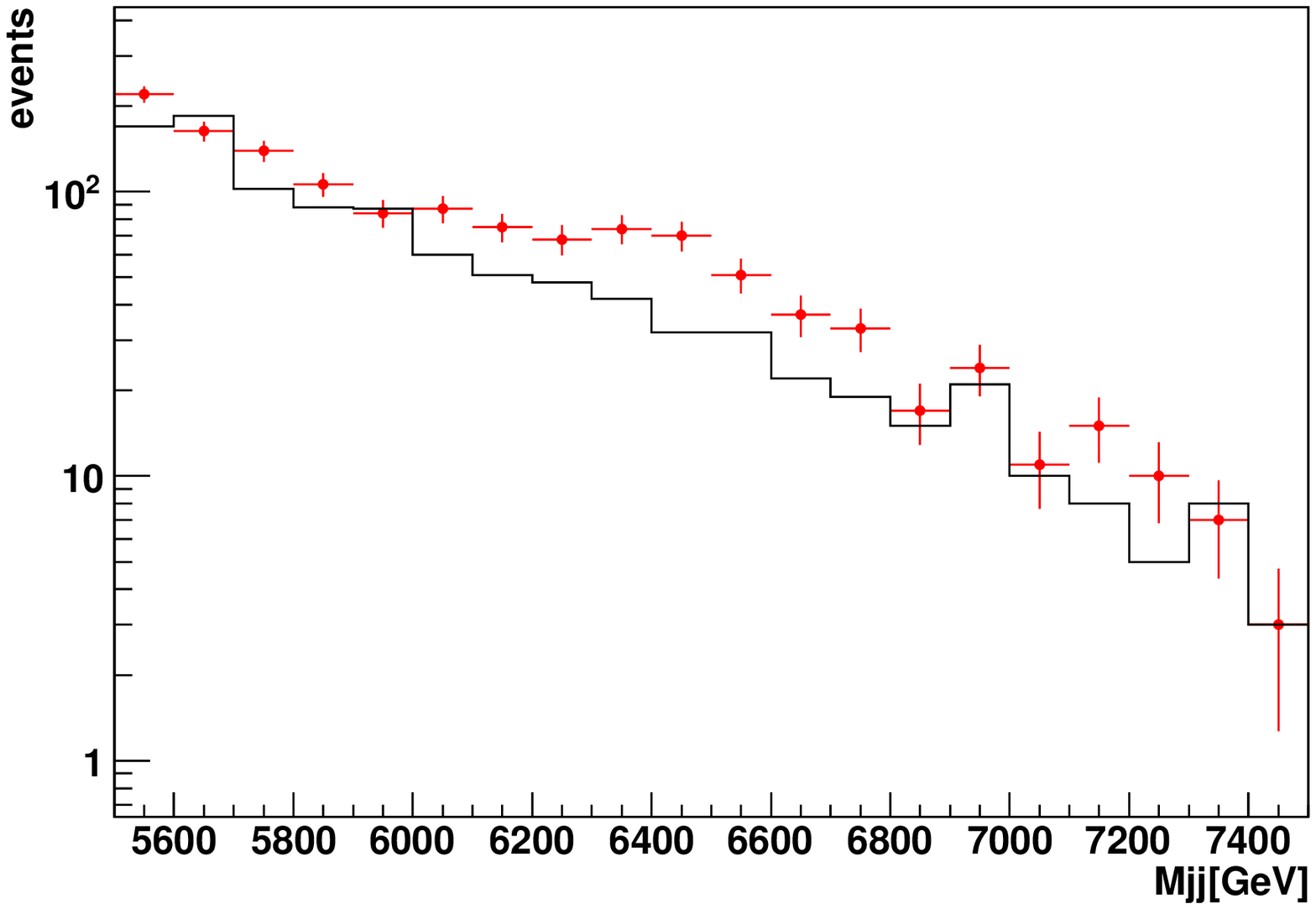}
	\end{minipage}
	\begin{minipage}{0.325\hsize}
		\centering
		\includegraphics[width=62mm]{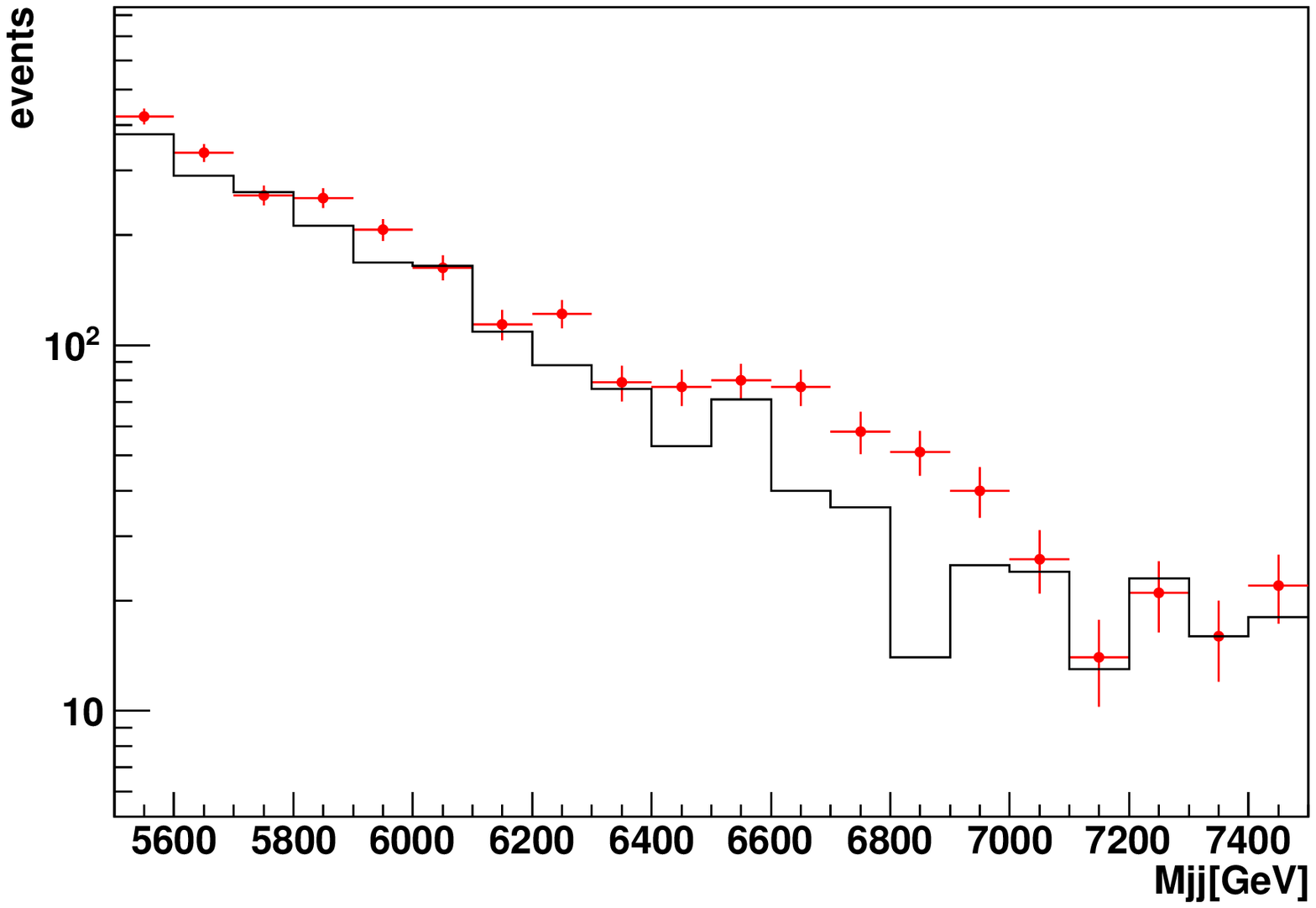}
	\end{minipage}
	\begin{minipage}{0.325\hsize}
		\centering
		\includegraphics[width=62mm]{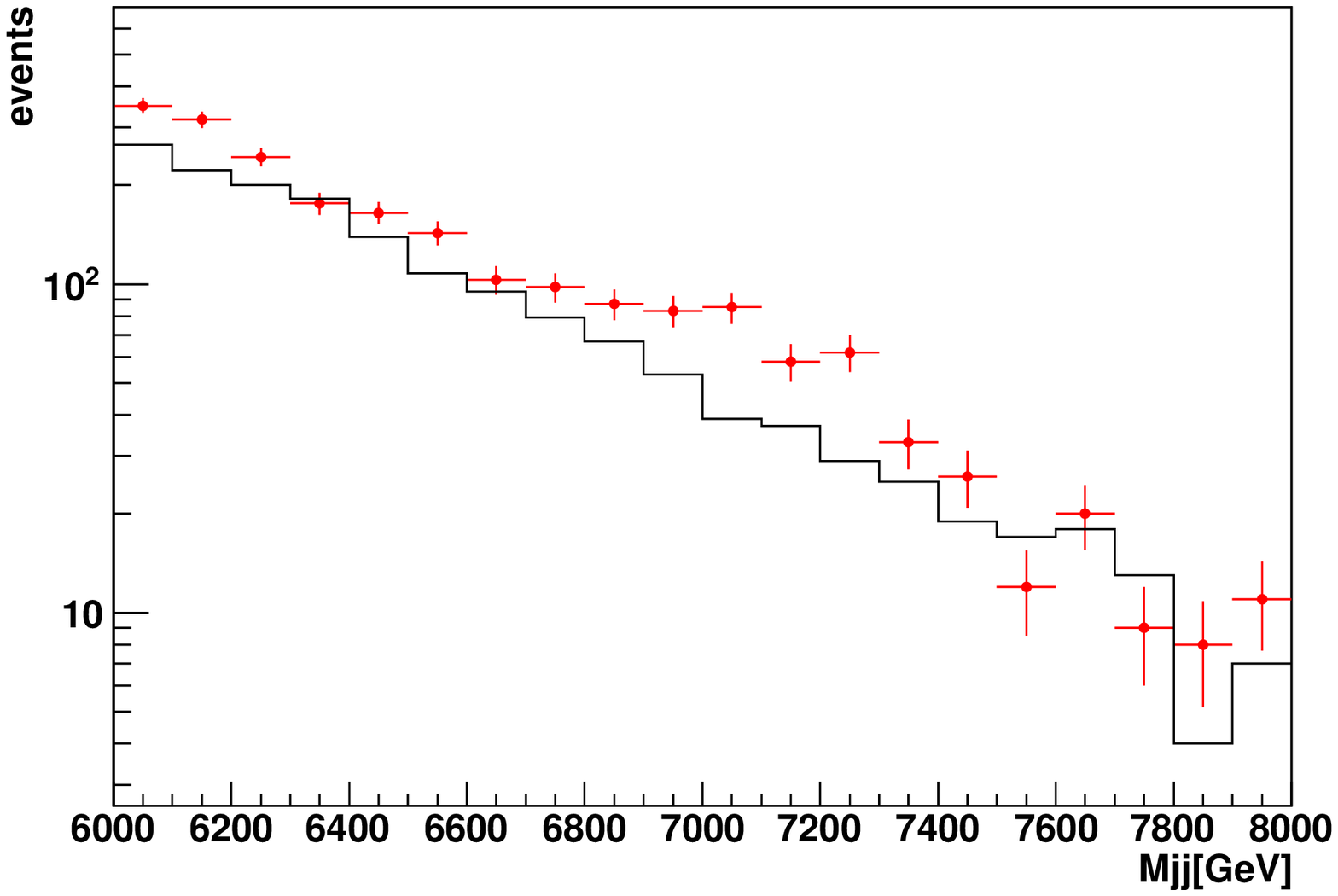}
	\end{minipage}
	\caption{\small{The dijet invariant mass distributions of MC data 
	with the second string resonances with the narrow widths (the points with error bars) 
	and of the SM background (the histogram), 
	in which the resonances are observed at $5\,\sigma$ level 
	for $M_\mathrm{s}=4.5\,\mathrm{TeV}$ with $150\,\mathrm{fb}^{-1}$ (the left figure), 
	$M_\mathrm{s}=4.75\,\mathrm{TeV}$ with $300\,\mathrm{fb}^{-1}$ (the middle figure) 
	and $M_\mathrm{s}=5\,\mathrm{TeV}$ with $600\,\mathrm{fb}^{-1}$ (the right figure).}}
	\label{fig:second_resonance}
	\end{figure}
\end{center}

Required integrated luminosities for $Z>5$ are $150$, $300$ and $600\,\mathrm{fb}^{-1}$ 
 for $M_\mathrm{s}=4.5$, $4.75$ and $5\,\mathrm{TeV}$, respectively, 
 if it is assumed that second string excited states do not decay into first string excited states.
The figures of Fig.\ref{fig:second_resonance} show the dijet invariant mass distributions 
 in which the second string resonances with the narrow widths of eq.(\ref{eq:2nd_quark_width}) 
 are observed at $5\,\sigma$ level.
We can see resonances at $M_{jj}=\sqrt{2}\times M_\mathrm{s}\simeq6.36$, $6.72$ and $7.07\,\mathrm{TeV}$ 
 for $M_\mathrm{s}=4.5$, $4.75$ and $5\,\mathrm{TeV}$, respectively.

\vspace{-14mm}
In the decay widths of $q^{**}$s in eq.(\ref{eq:2nd_quark_width}), 
 only a decay process of $q^{**}\rightarrow qg$ is considered.
However, decay channels into first string excited states, 
 namely decay processes of $q^{**}\rightarrow q^*g$ and $q^{**}\rightarrow qg^*$ should also be considered.
In order to avoid technical complication of 
 calculating decay widths of $\mathrm{2nd}\rightarrow\mathrm{1st}+\mathrm{SM}$, 
 we make a simple estimate of the widths to be the decay widths of $\mathrm{2nd}\rightarrow\mathrm{SM}+\mathrm{SM}$ 
 multiplied by a factor, following in Ref.\cite{Dong:2010jt}.
In other words, we describe total decay widths of $q^{**}$s as
\begin{equation}
	\begin{split}
		\label{eq:total_width}
	\Gamma_{q^{**},\,\mathrm{tot.}} 
	& = \Gamma_{q^{**}\rightarrow qg}
	+ \Gamma_{q^{**}\rightarrow q^*g}
	+ \Gamma_{q^{**}\rightarrow qg^*} \\
	& \equiv\Gamma_{q^{**}\rightarrow qg}
	+ a\times\Gamma_{q^{**}\rightarrow qg} \,,
	\end{split}
\end{equation}
 where $a$ is a constant factor.

The factor $a$ in eq.(\ref{eq:total_width}) is estimated in the following way.
First, we count the number of possible first string excited states of decay products.
The first string excited states of quarks, $q^*$s, have two states with $J=1/2$ and $3/2$ 
 and the first string excited states of gluons, $g^*$s, have three states with $J=0$, $1$ and $2$.
Therefore, the number of $q^*$s and $g^*$s of decay products is five.
Secondly, we consider phase space suppression.
A measure of the phase space in a decay process of 
 a state with mass $M$ into a massive state with mass $m$ and a massless state 
 is proportional to $(M^2-m^2)/2M^2$.
The phase space suppression factor between decay modes of 
 $q^{**}\rightarrow q^*g$ and $qg^*$ and $q^{**}\rightarrow qg$, 
 for example, can be estimated as 
 $\frac{\{(\sqrt{2}M_\mathrm{s})^2-M_\mathrm{s}^2\}/2(\sqrt{2}M_\mathrm{s})^2}{(\sqrt{2}M_\mathrm{s})^2/2(\sqrt{2}M_\mathrm{s})^2}=\frac{1}{2}$.
In total, the factor $a$ is estimated as $5/2$.

\begin{center}
	\begin{figure}[t]
	\begin{minipage}{0.5\hsize}
		\centering
		\includegraphics[width=62mm]
		{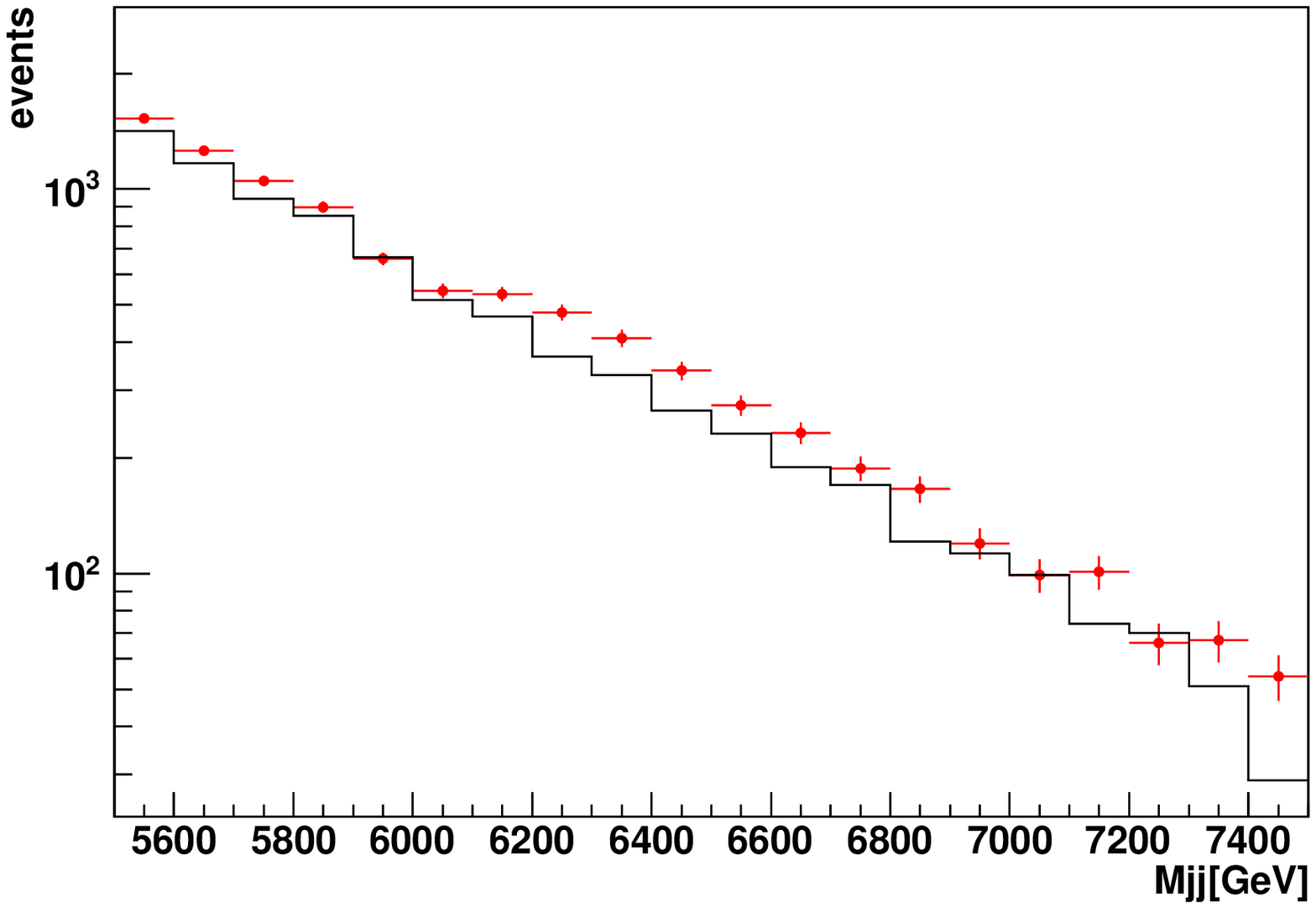}
	\end{minipage}
	\begin{minipage}{0.5\hsize}
		\centering
		\includegraphics[width=62mm]
		{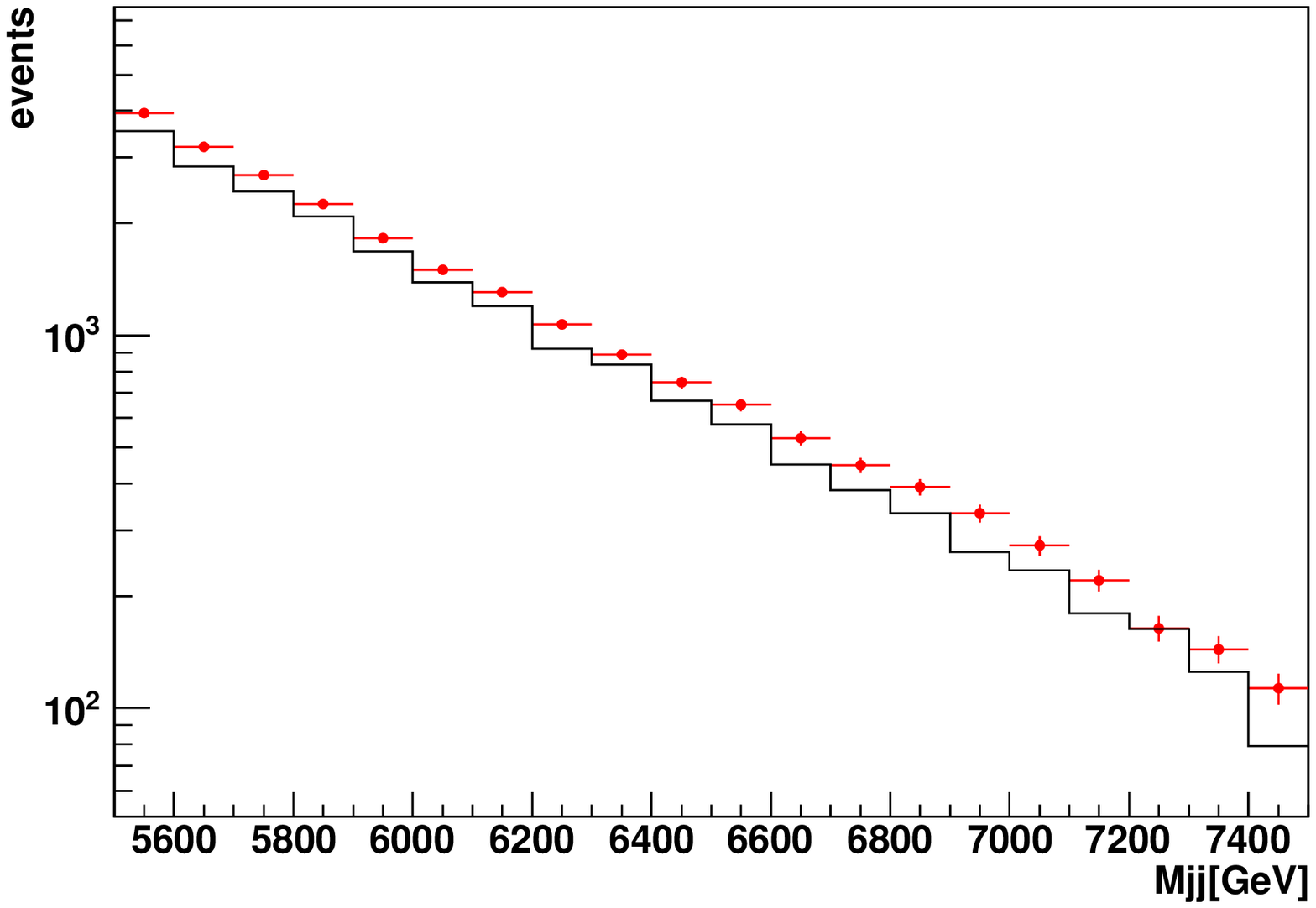}
	\end{minipage}
	\caption{\small{The dijet invariant mass distributions of MC data 
	with the second string resonances with the wider widths (the points with error bars) 
	and of the SM background (the histogram), 
	in which the resonances are observed at $5\,\sigma$ and $4.1\,\sigma$ level, 
	for $M_\mathrm{s}=4.5\,\mathrm{TeV}$ with $1200\,\mathrm{fb}^{-1}$ (the left figure) 
	and $M_\mathrm{s}=4.75\,\mathrm{TeV}$ with $3000\,\mathrm{fb}^{-1}$ (the right one), respectively.}}
	\label{fig:second_modified_width}
	\end{figure}
\end{center}

Event samples for the string signal are generated 
 in case of the improved widths of second string excited states,
 where the widths are $7/2$ times as large as that in eq.(\ref{eq:2nd_quark_width}).
The dijet invariant mass distributions with the second string resonances with the wider widths 
 are shown in Fig.\ref{fig:second_modified_width}, 
 for $M_\mathrm{s}=4.5\,\mathrm{TeV}$ and $4.75\,\mathrm{TeV}$, 
 assuming that $1200\,\mathrm{fb}^{-1}$ and $3000\,\mathrm{fb}^{-1}$ of integrated luminosities are given, respectively.
Within the $M_{jj}$ window of 
 $[\sqrt{2}M_\mathrm{s}-250\,\mathrm{GeV},\,\sqrt{2}M_\mathrm{s}+250\,\mathrm{GeV}]$, 
 we calculate significances are $Z=5$ for $M_\mathrm{s}=4.5\,\mathrm{TeV}$ with $1200\,\mathrm{fb}^{-1}$, 
 and $Z=4.1$ for $M_\mathrm{s}=4.75\,\mathrm{TeV}$ with even $3000\,\mathrm{fb}^{-1}$.

\vspace{-14.5mm}
In Fig.\ref{fig:second_modified_width}, the resonances almost do not form peaks.
The height of the resonance becomes much lower as the width and mass are larger even at parton-level.
In addition, the low-mass region of the resonance becomes wider 
 due to the effect of final state radiations and parton distribution functions.
On the other hand, the high-mass region of it also becomes wider 
 due to initial state radiations, 
 though the effect is not included in our MC simulations.

If second string excited states do not decay into first string excited states, 
 the second string resonance in dijet invariant mass distributions 
 can be discovered at $5\,\sigma$ level 
 with $\mathcal{O}(100)\,\mathrm{fb}^{-1}$ of integrated luminosity up to $M_\mathrm{s}=5\,\mathrm{TeV}$.
However, if they have decay channels of $\mathrm{2nd}\rightarrow\mathrm{1st}+\mathrm{SM}$, 
 larger than $\mathcal{O}(1000)\,\mathrm{fb}^{-1}$ of integrated luminosity is required 
 to discover the second string resonance at $5\,\sigma$ level.

\section{Conclusions}
	\label{sec:conclusion}

In low-scale string models, string excited states can be observed as resonances
 in dijet invariant mass distributions at the LHC,
 independently on the detail of the model buildings.
To distinguish low-scale string models from the other ``new physics'', 
 we have performed two analyses using MC simulations for the $14\,\mathrm{TeV}$ LHC.

The angular distribution analysis of dijet events is important, 
 because the degeneracy of string excited states with higher spins 
 is a distinct property of low-scale string models.
We have shown that contributions of several different spin states to a resonance 
 could be confirmed with $30\,\mathrm{fb}^{-1}$ and $50\,\mathrm{fb}^{-1}$ of integrated luminosities 
 for $M_\mathrm{s}=4.5\,\mathrm{TeV}$ and $5\,\mathrm{TeV}$, respectively.

The dijet invariant mass distribution analysis for the second string resonance is also important, 
 because the existence of second string excited states 
 with $\sqrt{2}$ times masses of that of first string excited states 
 is also a distinct property of low-scale string models.
We have shown that the second string resonance in dijet invariant mass distributions 
 could be discovered at $5\,\sigma$ level 
 with $150$, $300$ and $600\,\mathrm{fb}^{-1}$ for $M_\mathrm{s}=4.5$, $4.75$ and $5\,\mathrm{TeV}$, respectively.
These are results assuming that second string excited states do not decay into first string excited states.

In case of larger decay widths of second string excited states 
 including decay channels into first string excited states,
 we found that more integrated luminosities of $1200\,\mathrm{fb}^{-1}$ and larger than $3000\,\mathrm{fb}^{-1}$ 
 were required to discover the second string resonance at $5\,\sigma$ level, 
 for $M_\mathrm{s}=4.5\,\mathrm{TeV}$ and $4.75\,\mathrm{TeV}$, respectively.
This means that we might need the High-Luminosity LHC which is a possible future plan of the extension of the LHC.

\section*{Acknowledgments}

The authors would like to thank Takuya Kakuda for valuable comments.
N.K. is supported in part by Grant-in-Aid for Scientific Research on Innovative Areas ($\#2410505$) 
 from the Ministry of Education, Culture, Sports, Science and Technology of Japan.

\end{document}